\documentclass[aps,twocolumn,prd,superscriptaddress,preprintnumbers,showpacs]{revtex4}
\usepackage{graphicx}
\usepackage{bm}
\usepackage{amsmath}
\usepackage{amssymb}
\usepackage{amsthm}

\newcommand{\be}{\begin{equation}}
\newcommand{\ee}{\end{equation}}
\newcommand{\bea}{\begin{eqnarray}}
\newcommand{\eea}{\end{eqnarray}}
\newcommand{\bdm}{\begin{displaymath}}
\newcommand{\edm}{\end{displaymath}}
\newcommand{\beas}{\begin{eqnarray*}}
\newcommand{\eeas}{\end{eqnarray*}}
\newcommand{\M}{M}

\newcommand{\rv}{r_{\rm vir}}

\newcommand{\zf}{z_{\rm f}}
\newcommand{\zi}{z_{\rm i}}

\begin{document}

\title{Mass freezing in growing neutrino quintessence}

\author{Nelson J. Nunes}
\email[]{n.nunes@thphys.uni-heidelberg.de}
\affiliation{Institut f\"ur Theoretische Physik, Universit\"at Heidelberg, Philosophenweg 16, 69120 Heidelberg, Deutschland}
\author{Lily Schrempp}
\email[]{l.schrempp@thphys.uni-heidelberg.de}
\affiliation{Institut f\"ur Theoretische Physik, Universit\"at Heidelberg, Philosophenweg 16, 69120 Heidelberg, Deutschland}
\author{Christof Wetterich}
\email[]{c.wetterich@thphys.uni-heidelberg.de}
\affiliation{Institut f\"ur Theoretische Physik, Universit\"at Heidelberg, Philosophenweg 16, 69120 Heidelberg, Deutschland}

\date{\today}

\begin{abstract}
Growing neutrino quintessence solves the coincidence problem for dark energy by a growing cosmological value of the neutrino mass which emerges from a cosmon-neutrino interaction stronger than gravity. The cosmon-mediated attraction between neutrinos induces the formation of large scale neutrino lumps in a recent cosmological epoch. We argue that the non-linearities in the cosmon field equations stop the further increase of the neutrino mass within sufficiently dense and large lumps. As a result, we find the neutrino induced gravitational potential to be substantially reduced when compared to linear extrapolations. We furthermore demonstrate that inside a lump the possible time variation of fundamental constants is much smaller than their cosmological evolution. This feature may reconcile current geophysical bounds with claimed cosmological variations of the fine structure constant.
\end{abstract}


\maketitle

\section{Introduction}
\label{introduction}

The theoretical and observational implications of a coupling between a light quintessence field or cosmon and other matter species have been widely explored in the literature \cite{Wetterich:1987fk,Damour:1994zq,Wetterich:1994bg,Amendola:1999er,Holden:1999hm,Olive:2001vz,Nusser:2004qu,Nunes:2004wn,Koivisto:2005nr,Ringwald:2006ks,Schrempp:2006mk,Kesden:2006vz,Ahlers:2007st,Bean:2007ny,CalderaCabral:2009ja}. However, both cosmological tests and laboratory measurements challenge many of these scenarios and lead to strong bounds on the allowed strength of the coupling. 
This motivates the search for mechanisms that tend to naturally suppress the observable traces of these interactions which could be manifested, for example, in deviations from the gravitational force law or in the time-dependence of fundamental constants.

One possible suppression mechanism relies on the density dependence of fundamental couplings due to the density dependence of the value of a light scalar field \cite{Ellis:1989as}. If the mass of the scalar field is larger than the inverse size of a dense object, then the local value of the field typically adapts to the local density of the object 
 rather than to the cosmological evolution. For the example of the chameleon mechanism \cite{Khoury:2003aq,Khoury:2003rn,Brax:2004qh,Brax:2004px} the mass of the scalar field itself 
increases with density such that for sufficiently large coupling and dense objects the scalar field value decouples from the cosmological 
evolution. The scalar field, called a {\it chameleon field} under these circumstances, is allowed to interact with all forms of matter. 
The strength of the coupling  can be substantially larger than that of gravity.  Strong enough non-linearities are a general ingredient  for this type of mechanism to work.

Growing neutrino quintessence \cite{Amendola:2007yx,Wetterich:2007kr} is also characterized by important non-linearities. The coupling between the cosmon and neutrinos is substantially 
larger than the gravitational coupling and results in the formation of very large neutrino structures at late times \cite{Wintergerst:2009fh}. In this work, we focus on this scenario and discuss the implications for the gravitational potential of such neutrino lumps and on the variation of the electromagnetic fine structure constant in the presence of a field coupling to electromagnetism.

We find a strong {\it backreaction} effect which substantially reduces the neutrino mass within such neutrino lumps as compared to its 
cosmological value. In consequence, this effect leads to a substantial reduction of the neutrino-lump induced gravitational potential. 
Inside a neutrino lump the possible time variation of fundamental constants is strongly suppressed as compared to their 
cosmological time evolution. 

Strong non-linear features are present in this scenario despite the fact that the density dependent mass of the scalar field typically remains smaller than the inverse size of the object and that the chameleon mechanism does not operate.
The main ingredient of the non-linear behavior is the non-linear dependence of the local neutrino mass on the local value of the cosmon field. This combines with a non-linear cosmon potential. 

We assume that a very rapid increase of a local fluctuation in the neutrino number density ends after some virialization
process in a quasi-static lump containing a given number of neutrinos. 
For subsequent time we take a static neutrino number density $n_\nu(\vec{r})$ which is a function of the distance to the center of the lump. We then solve the field equation for the cosmon field $\phi$ taking into account the cosmological time variation of $\phi$ outside the lump.
The field equation for the local cosmon field $\phi$
has a source term proportional to $\beta m_\nu(\phi)n_\nu(\vec{r})$, where $\beta$ is the cosmon-neutrino coupling and may depend on $\phi$. The value of the neutrino mass $m_\nu(\phi)$ depends on $\phi$  and we assume, for simplicity, an equal mass for all three neutrino species. In the scalar field equation, the source term may be partly balanced by gradient terms and the derivative of the cosmon potential, $\partial V(\phi)/\partial \phi$. This  results in an
almost static value of the cosmon field which depends, however, on the neutrino density $n_\nu(\vec{r})$. 

A key ingredient of the growing neutrino quintessence scenario is the coupling of the  cosmon to the trace of the stress energy tensor of neutrinos, $(\rho_\nu-3p_\nu)$, where $\rho_\nu$ and $p_\nu$ denote the neutrino energy density and pressure, respectively. For most of the cosmological evolution neutrinos are relativistic and almost massless, thus $p_\nu \simeq \rho_\nu/3 $. Consequently, in this regime, the coupling is ineffective and both components (cosmon and neutrinos) evolve independently. The relativistic neutrinos permeate the universe homogeneously, only subject to cosmic expansion as in ``standard cosmology''. Furthermore, the cosmon tracks the background evolution following an attractor solution on trajectories characteristic of the presence of an exponential field potential~\cite{Wetterich:1987fm,Ratra:1987rm,Wetterich:1994bg,Copeland:1997et,Ferreira:1997hj}. However, as soon as the neutrinos turn non-relativistic, the interaction switches on as $p_\nu \simeq 0$. 
Since the  abundance of cosmic neutrinos is predicted by standard cosmology, the moment in time when neutrinos become non-relativistic is determined by  the present value of the mass as well as the strength of the cosmon-neutrino coupling and the cosmon potential. 
Typically, in the growing neutrino scenario, the cosmon-neutrino coupling exceeds the strength of the gravitational interaction by two to three orders of magnitude. The neutrino mass grows more rapidly for larger values of the coupling, but starts to increase later in time such that neutrinos become non-relativistic at redshift $z_{\rm NR} \sim 5-10$. 

As the neutrino mass grows, in turn, the neutrinos gain influence on the dynamics of the scalar field. Eventually the cosmon is forced out of the attractor and the early cosmological solution crosses over to an almost constant asymptotic field value at late times. Consequently, the energy density corresponding to the value of the exponential cosmon potential at the transition, acts similarly to a cosmological constant and drives cosmic acceleration. In the growing neutrino scenario, the corresponding dark energy scale is directly related to the particle physics scale of the present neutrino mass~\cite{Amendola:2007yx,Wetterich:2007kr}. This is a possible solution of the "why now?" problem
of dark energy.

Only after neutrinos have turned non-relativistic, the cosmon becomes the mediator of a ``fifth force'' between neutrinos. The strength of this interaction is proportional to the square of the coupling. 
The range of this force is of the order of the Hubble length, since the time evolving mass of the cosmon roughly equals the Hubble parameter. This is in contrast to adiabatic models of mass varying neutrinos (MaVaNs), which assume 
a scalar field with a mass much larger than the Hubble scale~\cite{Fardon:2003eh,Afshordi:2005ym,Brookfield:2005td,Brookfield:2005bz,Bjaelde:2007ki}. The mass of the cosmon emerges naturally from possible explanations of an exponential potential stemming from an asymptotically vanishing dilatation anomaly~\cite{Wetterich:2008sx}.

Due to the strong long-ranged force, at $z<z_{\rm NR}$, fluctuations in the non-relativistic neutrinos grow rapidly on all scales larger than the free-streaming length. Recent studies have widely explored the evolution of the growing neutrino perturbations both in the linear~\cite{Mota:2008nj,Pettorino:2009vn,Wetterich:2009qf} and in the non-linear regime~\cite{Wintergerst:2009fh}, finding a substantial clustering of neutrinos up to supercluster scales and beyond. Typically, neutrino perturbations with sizes larger than $\gtrsim 10$ Mpc become nonlinear at a redshift of $z\approx 1-2$ ~\cite{Mota:2008nj,Pettorino:2009vn,Wetterich:2009qf,Wintergerst:2010ui,Pettorino:2010bv}. Around this redshift the universe is thus populated by virialized neutrino lumps bound by the ``fifth force'' and by gravity. 
Present findings indicate that the neutrino profile within these lumps is consistent with a  highly-concentrated ``Navarro-Frenk-White'' (NFW) distribution.

Noting that the impact of these overdensities on the local scalar field dynamics is amplified by the large coupling, one can ask whether neutrino lumps provide a new source for highly non-linear effects.
In this paper we explore this question and analyze the resulting theoretical and observational consequences. To this end, we first determine the spatial cosmon profile within individual virialized neutrino lumps embedded within a time-evolving homogeneous neutrino background. 

By considering  example cases, we find the non-linear structure of the field equation to have the stunning effect of decoupling the value of the cosmon within a lump from its large-scale cosmological behavior. 
This implies a substantial reduction of the neutrino mass within neutrino lumps. To probe the dependence of these effects on the properties 
 of the  neutrino lumps, we investigate a family of 
neutrino distribution functions $n_\nu(r)$. For a given $n_\nu(r)$ we numerically solve the scalar field equation, with the boundary condition that outside the object the cosmon asymptotes to the cosmological value.  This allows us  
to determine the average neutrino mass in the interior of various families of neutrino lumps. The average neutrino mass turns out to decrease with increasing number of neutrinos in the lump. This implies a substantial reduction of the gravitational potential $\Phi_\nu$ of large scale neutrino lumps compared to linear extrapolations which are based on the cosmological neutrino mass. These findings reflect the importance of backreaction effects in the growing neutrino scenario. Furthermore, they bear significant consequences for the effect of neutrino lumps on the large scale structure and the cosmic neutrino background in the form of the Integrated Sachs-Wolfe effect \cite{Pettorino:2010bv,Brouzakis:2010md}. 

Finally, one may wonder whether the described non-linear effects in growing neutrino cosmologies might imply new interpretations of cosmological observations. For example, we consider the implications of our findings for the variation of the fine structure constant $\alpha$ which would result from a coupling of the cosmon to electromagnetism. We demonstrate that within large neutrino lumps, the non-linear effects suppress the time variation of the fine structure constant, while allowing for a strong variation in the background. This may reconcile geophysical bounds with claimed cosmological observations of a time variation of $\alpha$.

This paper is organized as follows. After setting the stage for our analysis, in Sec.~\ref{scenario} we set up the general equations governing the cosmon dynamics inside and outside of a neutrino lump.
In Sec.~\ref{numphi} we present our numerical results for the spatial cosmon profile and the neutrino mass and comment on the difference in field behavior compared to linear approximations.  In Sec.~\ref{grav}, we present our results for the average neutrino mass for various families of lumps and comment on the implications for their gravitational potential. We also discuss possible backreaction effects on the background cosmological value of the scalar field and the growth of large scale neutrino perturbations. In Sec.~\ref{varalpha} we explore the consequences of our finding for the variation of the fine structure constant in the presence of a scalar field coupled to electromagnetism.
Finally, in Sec.\ref{conc} we summarize our results and conclude. In a brief appendix, we derive relations between the parameters characterizing the NFW neutrino distribution function.

\section{Setting the stage}
\label{scenario}

In this work, it is our aim to estimate the spatial cosmon profile at late times for various families of neutrino lumps, formed at a redshift $z_{\rm NL} \simeq 1-2$.
To this aim we consider the perturbation to the field generated by a single virialized neutrino lump embedded in a time-evolving neutrino background of uniform density. This approximation is justified in the limit that the distance between lumps is much larger than their virial radius $\rv$.

As we will show in Sec.~\ref{numphi}, for a given static neutrino distribution in the interior and a given time-evolving background density well outside, the properties of a lump and its impact on the field can essentially be quantified by a single parameter, namely, the total number of neutrinos $N_\nu$ it contains.  
In this section, we will present the main ingredients of the growing neutrino scenario~\cite{Amendola:2007yx,Wetterich:2007kr} and describe the background evolution of the cosmon. Furthermore, we will set up the field equation for the spatial dependence of the cosmon
within the approach outlined above.  

\subsection{The growing neutrino scenario}
\label{gns}

We take both the scalar field potential $V(\phi)$ as well as the neutrino mass $m_{\nu}(\phi)$ to depend exponentially on the scalar field value,
\bea
V(\phi)=M^4\, \exp\left(-\alpha \frac{\phi}{\M}\right) \label{eq:Vexp}\, ,\\
m_{\nu}(\phi)=\hat{m}\, \exp\left(-\beta \frac{\phi}{\M}\right) \label{eq:mexp}\, ,
\eea
where $\M=(8\pi G_N)^{-1/2}$ denotes the reduced Planck mass, $\hat{m}$ is a constant that fixes the neutrino mass scale and $\alpha>0$ and $\beta<0$ are free model parameters. To comply with observational bounds on the amount of early dark energy, the parameter specifying the exponential potential is constrained to be $\alpha\gtrsim 10$
\cite{Doran:2007ep}. Note that according to our choice in Eq.~(\ref{eq:mexp}), the dimensionless coupling $\beta$ generally defined as
\be
\beta \equiv -M \frac{d\ln m_\nu}{d\phi} \,
\ee
is a constant. However, similar effects to the ones we encounter for a constant coupling are expected in the case of a field dependent coupling, $\beta(\phi)$, as has been suggested within a particle physics context~\cite{Wetterich:2007kr}.

As in previous works~\cite{Amendola:2007yx,Wintergerst:2009fh}, we assume $\alpha = 10$ and $\beta = -52$ for our numerical studies. 
We choose a present cosmological neutrino mass $\bar{m}_{\nu0}= 2.3$eV
consistent with current cosmological and experimental bounds.
The value of $\hat{m}$ in Eq.~(\ref{eq:mexp}) is then dictated by the precise timing of the transition from matter to cosmon domination according to the observed dark energy abundance $\Omega_{\rm de} \approx 0.73$. 

\subsection{Time evolution of the cosmon in the cosmological background}\label{time}
The dynamics of the cosmological background field $\bar{\phi}$ is described by the Klein Gordon equation,
\be
\ddot{\overline{\phi}}+2\mathcal{H} \dot{\overline{\phi}} +a^2\frac{d V}{d\overline{\phi}}=a^2\frac{\beta}{M}(\rho_\nu(\overline{\phi})-3p_\nu(\overline{\phi})),\label{eomtime}
\ee
where dots denote the derivative with respect to conformal time $\tau$ and $\mathcal{H} =\dot{a}/a$ is the corresponding Hubble parameter. Furthermore, $\rho_\nu$ and $p_\nu$ denote the neutrino energy density and pressure, respectively. One observes that the neutrino source term on the right hand  side of Eq.~(\ref{eomtime}) is strongly suppressed as long as neutrinos are relativistic, since $p_\nu \simeq \rho_\nu/3$. However, when they turn non-relativistic,  $p_\nu\simeq 0$, the source term $\beta\rho_\nu/M$ becomes of dynamical importance. For constant $\beta$, the energy densities of the cosmon and of the neutrinos evolve asymptotically proportional to each other. Averaging over oscillations in time, the cosmological value $\bar{\phi}$ evolves as 
\begin{eqnarray}
\label{background}
\bar{\phi}(z) &=& \bar{\phi}_0 - \frac{3M}{\alpha-\beta}  \ln (1+z) \,, \\\nonumber
\bar{\phi}_0 &=& \frac{M}{\alpha} \ln \left( \frac{M^4}{\rho_{\rm de}} \left(1-\alpha/\beta\right)\right) \,,
\end{eqnarray}
with $z$ being the cosmic redshift and $\rho_{\rm de}=V(\phi)+\rho_{\nu}\simeq V(\phi)$ denoting the dark energy density.

\subsection{Quasi-static cosmon field in a neutrino lump}
\label{spatial}
Initial neutrino perturbations can evolve into virialized lumps with a highly concentrated NFW type neutrino distribution. Due to the neutrino-cosmon coupling, 
the spatial profile of the scalar field $\phi(r)$ resulting from a given static  spherically symmetric distribution $n_\nu(r)$ of non-relativistic neutrinos is determined from the following field equation:
\be
 \phi^{\prime \prime}+\frac{2}{r}\phi^{\prime}=\frac{{d}V(\phi)}{{d}\phi}-\frac{\beta}{M}\rho_\nu(\phi,r) \,,
\label{eom}
\ee
where
\be
\rho_{\nu}(\phi,r) = n_{\nu}(r) \, m_{\nu}(\phi) \,.
\label{rhonr}
\ee
Here and in what follows a prime denotes a derivative with respect to the radial coordinate 
$r$. Furthermore, the last equality applies for non-relativistic neutrinos, with $\rho_\nu$ and $n_\nu$ denoting the neutrino energy density and number density, respectively. 

In growing neutrino quintessence, the source term in the spatial field equation depends of the field value in contrast to other classes of coupled dark energy scenarios 
\cite{Schrempp:2009kn}.
The reason is that $m(\phi)$ in Eq.~(\ref{eq:mexp}) cannot be linearized as 
$\beta \phi/M \gg 1$ owing to the large coupling and the large field value. This issue will be discussed in detail in Sec.~\ref{numphi}.   

The solution of the field equation (\ref{eom}) has to obey the following
boundary conditions 
\be
\phi'(r)|_{r=0} = 0  \hspace{0.5cm} {\rm and} \hspace{0.5cm} \phi(r)|_{r\rightarrow \infty} = \bar{\phi} \,.\label{ini}
\ee
The first condition follows from requiring the field value to be finite and well defined at $r=0$ and the second implies that the field must converge to its cosmological background value $\bar{\phi}$ in the limit of $r$ going to infinity. 
The possible dependence of the solution on redshift is induced by the redshift dependence of the background field value $\bar{\phi} = \bar{\phi}(z)$. 

We assume here, implicitly, that the neutrino number distribution $n_\nu(\vec{r})$ for a virialized lump has decoupled from the cosmological evolution and therefore is time-independent. In addition to the static lump distribution $n_\nu(\vec{r})$, the cosmon also couples to the cosmological background neutrino density $\bar{n}_\nu(z)$ which is much smaller than $n_\nu(\vec{r})$ within the lump. This coupling to the background, therefore, only plays a role at the border or outside the lump.

\subsection{Neutrino distribution}
\label{neudis}

The spherically averaged radial distribution of neutrinos within virialized lumps is typically described by a highly-concentrated Navarro-Frenk-White profile~\cite{Wintergerst:2009fh}
\be
n_{\rm NFW}(r)=\frac{n_s}{(r/r_s)\left(1+r/r_s\right)^2} \,,
\label{NFW}
\ee
where $n_s$ is the normalization of the number density which can be different for different lumps.
The characteristic scale $r_s$ separates the inner part $r\ll r_s$ with $n_{\rm NFW}\propto r^{-1}$ from the outer region, where 
$n_{\rm NFW}\propto r^{-3}$. 
The number of neutrinos contained within a radius $R$ can be written as 
\be
N_{\nu}(R) = 4\pi\int^{R}_{0}{d}r \, r^2 \, n_{\rm NFW}(r) = N_* \, F(R)\,, \label{Nnfw}
\ee
with 
\bea
\label{nstar}
N_*&=&4\pi n_s r_s^3\,,\\
\label{fR}
F(R) &=&\ln\left(1+\frac{R}{r_s}\right)-\frac{R}{R+r_s} \,.
\eea
Because the increase of $N_\nu(R)$ for large $R\gg r_s$ is only logarithmic,
\be
N_\nu(R)\simeq N_* \ln\left(\frac{R}{e\,r_s}\right),
\ee
we can use $N_*$ as a reasonable parametrization for the number of neutrinos in the lump. More exactly, $N_*$ approximately gives the number of neutrinos within a radius
$10 r_s$. 

This implies that instead of $n_s$
and $r_s$ one may use $N_*$ and $r_s$ for the parametrization of the profile $n_\nu(r)=n_{\rm NFW}(r)$. We typically compare 
$N_*$ with the total number of neutrinos within our horizon, 
\be 
N_{\rm tot} = \frac{4\pi \bar{n}_{\nu 0}}{3\,H^3_0}\simeq 2.4\times 10^{87}\,, 
\ee
where $n_{\nu 0}$ denotes the present cosmological number density of neutrinos and $H_0$ the present Hubble parameter. Thus, for a NFW profile, we will characterize the neutrino lumps by the two numbers $N_*/N_{\rm tot}$ and $r_s$.

It is also possible that the logarithmic increase of $N_\nu(R)$ does not extend to arbitrary large $R$. This happens if a neutrino lump competes for the neutrinos in neighboring lumps. One may imagine that the infall of further neutrinos is stopped at some characteristic time and subsequently all available neutrinos concentrate in a finite volume of size characterized by a virial radius $\rv$. We account for this possibility by considering a second type of profile
\be
n_{\rm step}(r) = \begin{cases}
           n_{\rm NFW}(r) \,, & r <r_{\rm vir} \\
           0 \,, &    r > r_{\rm vir}
           \end{cases}
\label{n1}
\ee
In what follows we will refer to these two possibilities Eqs.~(\ref{NFW}) and (\ref{n1}) as the "pure NFW" and the "step NFW", respectively. We will see that the pure NFW (which corresponds to $\rv \rightarrow \infty$) and the step NFW yield qualitatively similar results.

The step NFW profile (\ref{n1}) implies a well specified number of neutrinos corresponding to the lump
\be
N_{\nu}=4\pi\int^{\rv}_{0}{d}r \, r^2 \, n_{\rm NFW}(r) = N_* \, F(c)\,,
\label{nonu}
\ee
with
\be
\label{Fc}
F(c)=\ln(1+c)-\frac{c}{1+c}\simeq \ln c -1\,,
\ee
where $c = r_{\rm vir}/r_s$ and the last equality holds for $c\gg 1$. A neutrino lump is in this case characterized by three parameters: $N_*/N_{\rm tot}$, $r_s$ and the concentration $c$. 
For a finite $\rv$ or $c$ we may also replace $N_*/N_{\rm tot}$ by $N_\nu/N_{\rm tot}$.

Note that for each class of lumps with given $N_{\nu}$ and $\rv$ the maximal value for the number of lumps $\mathcal{N}_l$ within the horizon is
\be
\mathcal{N}_l=\frac{N_{\rm tot}}{N_{\nu}}\label{maxlumps} \,.
\ee
This corresponds to the limit when all neutrinos in the visible universe are captured in identical lumps. 

In order to satisfy the second of the boundary conditions in Eq.~(\ref{ini}) we must add to the lump profile a cosmological background number density $\bar{n}_{\nu}(z)$ such that 
\begin{equation}
\label{numberdensity}
n_{\nu}(z,r)= n_{\rm NFW}(r) \, \theta(r_{\rm vir} -r) + \bar{n}_\nu(z) \,,
\end{equation}
with
\be
\bar{n}_\nu(z)=\bar{n}_{\nu 0}\,(1+z)^3\label{ninf} \,,
\ee 
where $\bar{n}_{\nu 0}$ corresponds to the present value 
and we recall that $\rv \rightarrow \infty$ for the pure NFW profile.
Of course Eq.~(\ref{ninf}) only remains a reasonable estimate for the neutrino number density provided most neutrinos are not yet concentrated in lumps. Once most of the neutrinos are in lumps, the time evolution of the cosmological average value of the quintessence field $\bar{\phi}(z)$ is more complicated and backreaction effects ought to be taken into account. For our purposes Eq.~(\ref{ninf}) is sufficient, since we are only interested in the effects of a given time evolution $\bar{\phi}(z)$ on the cosmon profile inside a lump and not in the computation of the evolution of $\bar{\phi}(z)$ itself. In general, the redshift dependence of $\bar{\phi}(z)$ can be made consistent by an approximate formal choice of $\bar{n}_\nu(z)$ in Eq.~(\ref{numberdensity}). 
 The detailed redshift dependence of $\bar{\phi}$ and $\bar{n}_\nu(z)$ is unimportant for the findings of this work.

\section{Mass freezing}
\label{numphi}
In this section we demonstrate that the neutrino mass inside of sufficiently dense and large neutrino lumps does not follow the cosmological evolution.
For this purpose, we numerically solve the cosmon field equation, Eq.~(\ref{eom}), for spherically symmetric static neutrino distributions (\ref{NFW}) and  (\ref{n1}). The boundary condition 
$\phi^{\prime}(r=0)=0$ ensures the regularity of the solution. The second boundary condition is given by the asymptotic value of $\phi$ for large $r \rightarrow \infty$, $\phi(r\rightarrow \infty)=\bar{\phi}$ (see Eq.~(\ref{ini})). The value $\bar{\phi}$ must equal the cosmological
average value of the cosmon field (or more precisely, the average over regions without lumps). Thus we compute the solutions for different "boundary values"
$\bar{\phi}$. The time or redshift dependence of the cosmon field in the lump is then induced by inserting the cosmological value of $\bar{\phi}(z)$. This procedure assumes that the adaptation of the cosmon field inside the lump to the cosmological value occurs on a sufficiently small time scale compared to the inverse Hubble parameter, and that the use of a static solution provides a good approximation. In practice we impose the boundary value 
$\bar{\phi}(z)$ for some very large value of $r \gg r_s$ where the derivative of the potential and the source term in the field equation nearly cancel each other.

Let us start by discussing the approximation of linear field equations, where the field  dependence of the neutrino mass $m_\nu(\phi)$ in Eq.~(6) is neglected and 
$\partial V/\partial \phi \approx 0$. In this case, the general
solution can be written as $\phi(r)=\bar{\phi}+\delta\phi(r)$, where $\delta\phi(r)$ vanishes for $r\rightarrow \infty$ and is independent of $\bar{\phi}$. 
A cosmological change of $\bar{\phi}$ would then result in a displacement of the cosmon field for the whole lump configuration. Such a linear
approximation would be valid for uncoupled quintessence or sufficiently small $\beta$ -- in this case the cosmon field acts as a "cosmological clock". If the variation of fundamental 
couplings (such as the fine structure constant) can be written as a linear dependence on the displacement of $\phi$, then, their evolution inside the lump follows the cosmological evolution~\cite{Wetterich:2002ic,Mota:2003tm,Shaw:2005vf}.

In our case of strong coupling, $\beta\gg1$, the linear approximation is not justified as the $\phi$-dependence of $m_\nu\propto \exp(-\beta\phi/M)$
plays an important role. We demonstrate this here for the case of a constant $\beta$. For a $\phi$-dependent $\beta(\phi)$ the effect may even 
be more pronounced. 
\begin{figure}
\includegraphics[width=8.5cm]{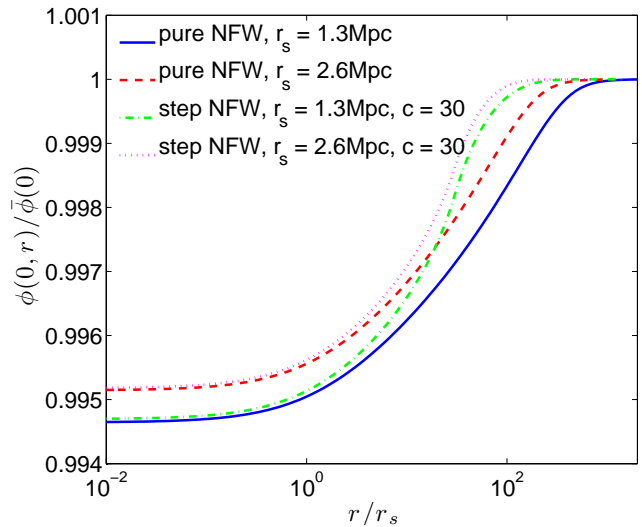}
\caption{\label{fig1} Scalar field profile $\phi(r,z=0)$ in the radial direction for a pure NFW neutrino profile (Eq.~(\ref{NFW}))
and for a step NFW profile (Eq.~(\ref{n1})). For both cases $N_* = 4\times 
10^{-3}N_{\rm tot}$. 
We show curves for $r_s = 1.3$Mpc and $r_s = 2.6$Mpc. 
For the step NFW profile we set $c = 30$.}
\end{figure}
In Fig.~\ref{fig1} we illustrate the radial dependence of $\phi(r)$ for the
pure and step NFW neutrino profiles for a lump with 
$N_*=4\times10^{-3}N_{\rm tot}$, evaluated 
for $\bar{\phi}_0 = \bar{\phi}(z=0)$ corresponding to the present cosmological value of 
the field. The values of the parameters $r_s$ will be motivated at the end of this section.
We find a qualitatively similar behavior $\phi(r)$ for the two profiles. In the outer region of the lump, the value of $r_s$ is not important. As one might have expected, the deviations of $\phi(r)$ from the cosmological value $\bar{\phi}$ occur for larger $r$ for the pure NFW profile.  In the inner region of the lump mainly $r_s$ matters (for a given $N_*$), while the value of 
$r_{\rm vir}$ only plays a minor role. Though the typical deviations of $\phi(r)$ from the cosmological value are only half a percent, their impact on the value of the neutrino mass is of considerable importance due to the large value of the coupling $\beta$.

For the various profiles of $n_\nu(r)$, we found the value of $m_\nu$
in the interior to be roughly three to four orders of magnitude smaller than the present cosmological one, $\bar{m}_{\nu 0}$. 
This significant reduction owes to the exponential dependence $m_\nu(r)/\bar{m}_{\nu}=  \exp(-\beta (\phi(r)-\bar{\phi})/M)$, 
which makes the neutrino mass very sensitive to the decrease of the field value in the interior of a lump. 
The implications of this result for the gravitational potential of a neutrino lump as well as for 
the cosmological evolution of the cosmon will be discussed in Sec.~\ref{grav}.

\begin{figure}
\includegraphics[width=8.5cm]{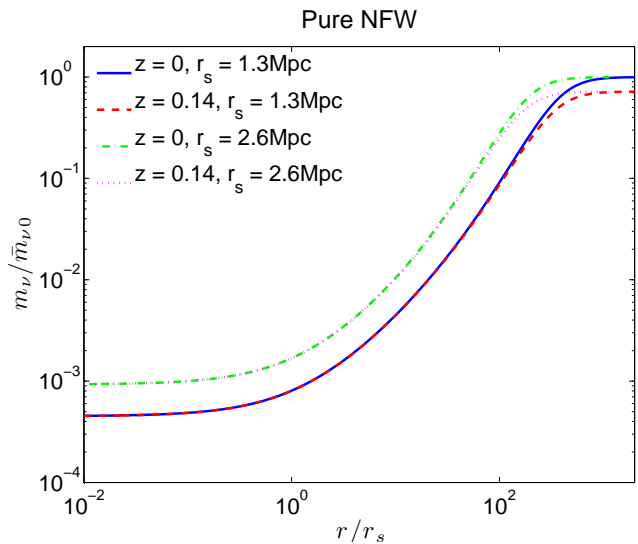}
\caption{\label{Masses1} Neutrino mass profile $m_\nu(r)/\bar{m}_{\nu 0}$ in the radial direction for redshifts $z = 0$ and 
$z = 0.14$ for the pure NFW profile.}
\end{figure}
\begin{figure}
\includegraphics[width=8.5cm]{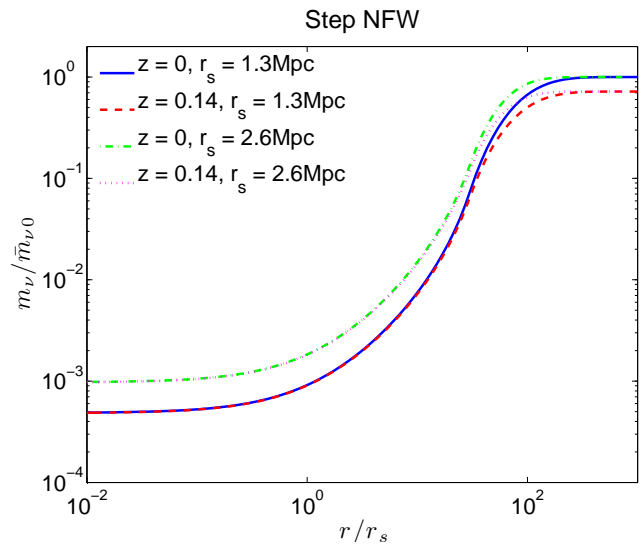}
\caption{\label{Masses2} Neutrino mass profile $m_\nu(r)/\bar{m}_{\nu 0}$ in the radial direction for redshifts $z = 0$ and 
$z = 0.14$ for the step NFW profile.}
\end{figure}
These findings suggest that the neutrino mass inside the lump does not follow the cosmological evolution. We demonstrate this in Figs.~\ref{Masses1} and \ref{Masses2}, where we 
evaluate the neutrino mass for the same neutrino profiles of Fig.~\ref{fig1} and 
for two different redshifts $z=0$ and $z=0.14$. 
We compare the neutrino mass as a function of the distance from the center of the lump with the present cosmological value of the neutrino mass, $\bar{m}_{\nu 0}$. For large $r$ the neutrino mass follows the cosmological evolution independently of the properties of the lump. In contrast, for small $r$ the change of the neutrino mass is very substantially reduced. The curves for the same
profile $n_\nu(r)$ but with different $\bar{\phi}(z)$ are nearly on top of each other. 
Inside the lump, the value of $m_\nu(r)$ depends essentially only on the parameters characterizing the neutrino number density profile of the lump, mainly on $N_*$ and to a smaller extent, on $r_s$.
This demonstrates very directly the "mass freezing" effect. The time evolution of the neutrino mass inside the lump is frozen at its value at some redshift characteristic for the formation of the lump. It depends on the neutrino density of the lump, but not on the cosmological evolution. This decoupling of the local properties from the cosmological evolution is very similar to the decoupling of the gravitational potential of a galaxy from the cosmological expansion. We will discuss in Sec.~\ref{varalpha} that this decoupling can have important consequences for a possible observation of time variation of fundamental constants. 

\begin{figure}
\includegraphics[width=8.5cm]{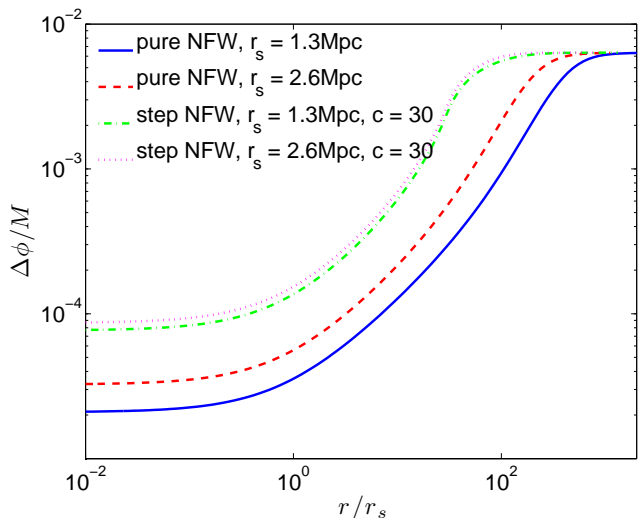}
\caption{\label{Shifts} Time change of the cosmon profile $\Delta\phi(r)=\phi(z=0,r)-\phi(z=0.14,r)$ in the radial direction for four neutrino number density profiles.}
\end{figure}

We demonstrate the decoupling explicitly in Fig.~\ref{Shifts}, where we show the difference of the local field between $z=0$ and $z=0.14$, 
$\Delta  \phi(r) =\phi(z=0,r)-\phi(z=0.14,r)$ for the neutrino number densities described above. 
The figure illustrates that in the considered example cases the radial field profile in the interior changes up to three orders of
 magnitude less than for large $r$. It can be verified that a larger $N_*$ enhances this effect.
 
\begin{figure}
\includegraphics[width=8.5cm]{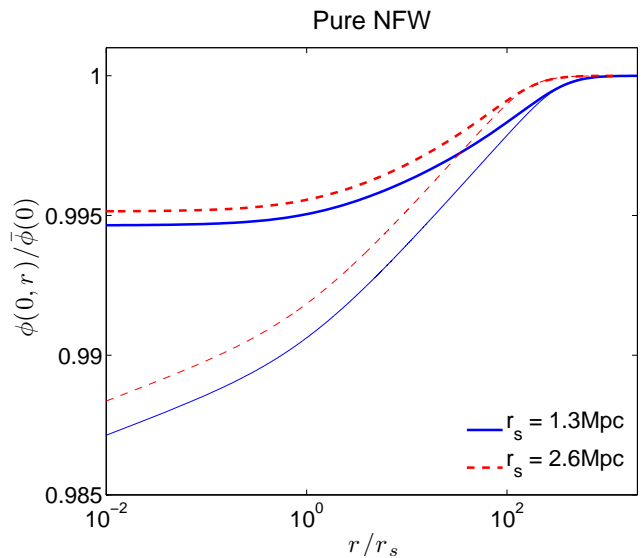}
\caption{\label{nogradpure} Comparison of the local density approximation (thin lines) with the numerical solution (thick lines) for the scalar field profile for a pure NFW neutrino profile.}
\end{figure}

\begin{figure}
\includegraphics[width=8.5cm]{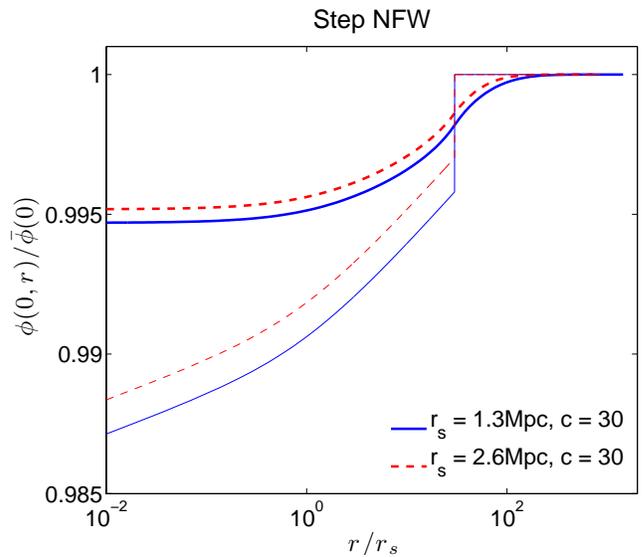}
\caption{\label{nogradstep} Comparison of the local density approximation (thin lines) with the numerical solution (thick lines) for the scalar field profile for a step NFW neutrino profile.}
\end{figure}

We conclude that non-linear effects turn out to be very strong and it is interesting to compare the behavior of the solution within the lump with a "local density approximation" 
where the gradient terms in the scalar field equation are neglected. In this case, the cosmon field locally follows the neutrino number density according to,
\be
\frac{\partial V(\phi)}{\partial \phi}=\frac{\beta}{M}m_\nu(\phi)n_\nu(r) \,,
\ee
or equivalently
\bea
\label{localdens}
\alpha M^3 e^{-\alpha\bar{\phi}_0/M}e^{-\alpha(\phi-\bar{\phi}_0)/M}= \hspace{2cm} \nonumber \\
\hspace{2cm} -\frac{\beta}{M}e^{-\beta(\phi-\bar{\phi}_0)/M} \bar{m}_{\nu 0} \, n_\nu(r) \,.
\eea
Here we choose a constant $\bar{\phi}_0$ and assume that $V(\bar{\phi}_0)$ equals the present dark energy density, 
\be
V(\bar{\phi}_0)\simeq 0.7\rho_c \simeq 10^{-120} M^4,
\ee
where $\rho_c$ denotes the present critical energy density. 
We compare in Figs.~\ref{nogradpure} and \ref{nogradstep} the numerical solution of Eq.~(\ref{eom}) with the local density approximation (\ref{localdens}) for various neutrino number profiles $n_\nu(r)$. We observe that though the local density approximation and the numerical solution are convergent for large $r$, they are fairly different for small $r$. This is a reflection of the fact that the cosmon mass is not large enough for the chameleon mechanism to operate in our growing neutrino quintessence model~\cite{Schrempp:2009kn}. 
In other words, the boundary conditions (\ref{ini}) prevent the field to sit at the 
minimum of the effective potential, $\phi_{\rm min}$, but instead at larger values. The field 
cannot be at smaller values $\phi(r=0) < \phi_{\rm min}$, because this would lead  
to a runaway solution (if we think of $r$ as the time variable) and the boundary 
condition at infinity would not be reached. For the local approximation to work, 
the mass of the cosmon would have to be much larger than the inverse size of the lump.

We finally comment on our choice of $r_s$ for a given value of $N_{*}$. Even though the dependence of our numerical results on $r_s$ is moderate, as can be seen by comparing Figs.~2 and 3, it is substantial enough to warrant some estimate of this parameter. For a given $N_{*}$, the radius $r_s$ determines the density in the core of the lump. Realizing that $n_{\rm NFW}(r_s)=n_s/4$, one has $n_{\rm NFW}(r_s)=N_{*}/16\pi r_s^3$. For $r\ll r_s$ one finds $n_{\rm NFW}(r)\simeq n_s r_s/r\simeq N_*/4\pi r^2_s r$. Thus smaller $r_s$ enhances the core density. This qualitative feature also applies to lumps with a finite size $r_{\rm vir}$, if we keep $N_{\nu}$ fixed (both for fixed $c$ or fixed $r_{\rm vir}$).

Some insight on orders of magnitude can be gained from the early stages of the formation of individual lumps which has been analyzed in Ref.~\cite{Wintergerst:2009fh}. Given that the cosmon-neutrino interaction is ineffective during the relativistic neutrino regime, the initial power spectrum for the linear neutrino perturbations is the same as in standard cosmology. Once the neutrinos turn non-relativistic, the cosmon-neutrino interaction becomes effective and, as a result of the large coupling, the neutrino perturbations grow rapidly on all scales larger than the free-streaming length. 

After a first phase, where an initial neutrino fluctuation accompanies the universe's expansion, its potential energy becomes important with respect to the kinetic energy of the neutrinos it contains. Eventually, the perturbation reaches its maximal size and thereafter, with increasing inward velocities, contracts according to the laws of free fall until virialization occurs and the collapse terminates. 

\begin{figure}
\includegraphics[width=8.5cm]{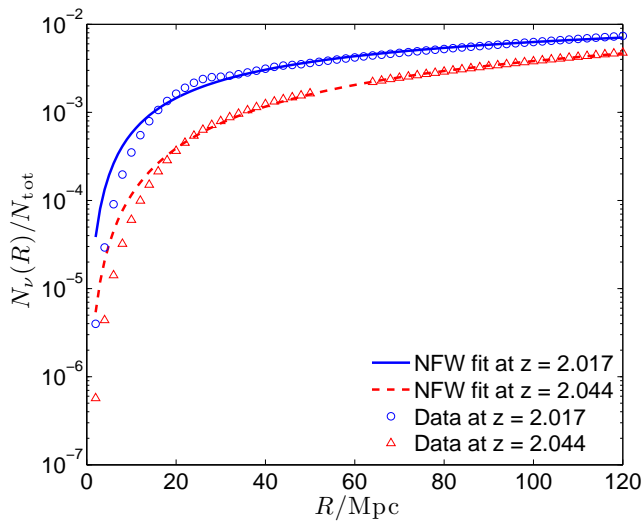}
\caption{\label{N} The number of neutrinos $N_\nu$ within a sphere of radius $R$ at two epochs during the collapse of a lump. The lines represent the corresponding fits to a pure NFW-profile. The data on 
$N_\nu$  were obtained using results of a dedicated numerical analysis.}
\end{figure}

The Navier-Stokes equations describe the formation process of an individual neutrino lump in an expanding universe in physical space as a function of time. A numerical code has been developed
\cite{Wintergerst:2009fh}, which allows one to track the corresponding time evolution of the relative neutrino density perturbation $\delta_\nu\equiv\delta\rho_\nu/\bar{\rho}_\nu$ as well as the cosmon perturbation $\delta \phi\equiv \phi(r)-\bar{\phi}$ from the linear regime up to the highly non-linear regime, until shortly before the neutrino lump virializes. In this code, there is no assumption on the final number density profile of the lump. The neutrino number density distribution is related to the spherically averaged density contrast $\delta_\nu$ and field perturbation $\delta\phi$ at a given redshift $\zf$ by,
\be
n_\nu(r,\zf)=\bar{n}_{\nu 0} (1+\zf)^3 (1+\delta_\nu(r,\zf))\, e^{\beta\delta\phi(r,\zf)/{\M}} \,.\label{nlump}
\ee
Using results of this numerical analysis, we show in Fig.~\ref{N} the number of neutrinos $N_\nu(R)$ within a sphere with radius $R$, 
\be
N_\nu(R)=4\pi\int^{R}_{0}d r\,r^2 n_\nu(r) \,,
\ee
at two moments in time: when the code terminates ($\zf=2.017$) and shortly before ($\zi=2.044$). (The termination of the code due to numerical instabilities can be associated with the onset of virialization. 
At this moment the potential and the kinetic energy of the neutrinos in the lump have reached roughly equal values.) We have fitted this particular case, which corresponds to a final gravitational radius $R_{\rm f}=72.1$Mpc, to the NFW profile (cf. Eq.~(\ref{Nnfw})). 
We define $R_{\rm f}$ as in  Ref.~\cite{Wintergerst:2009fh}, i.e. the distance to the center of the lump where the gravitational potential is a factor 2 smaller than the potential at the center.

For $\zi=2.044$ one finds $N_{*}/N_{\rm tot}=1.2\times10^{-2}$, $r_s=64.2$Mpc, while $\zf=2.017$ yields $N_{*}/N_{\rm tot}=5.5\times 10^{-3}$, $r_s=15.7$Mpc. It is clear that at these redshifts the neutrinos continue to fall into the structure, resulting in an increase of $N_{*}$ and $r_s$ with $z$. Nevertheless, this gives already an idea of the order of magnitude of these quantities for larger lumps. Once the further infall of neutrinos is stopped or substantially reduced, e.g. by competing lumps, we expect the concentration process of the neutrinos to go on for a while until a static virialized situation is reached. This concentration process leads to a reduction of $r_s$ for almost constant $N_{*}$. This motivates the values of $r_s$ used in our figures. For lumps with a finite radius $r_{\rm vir}$ we may also relate $n_s$ to the neutrino number density contrast $\Delta_n \equiv \langle n \rangle/\bar{n}_{\nu 0}$. This is discussed in the appendix.

\section{Gravitational potential of neutrino lumps}\label{grav}
The strength of the non-linear effects depends on the number of neutrinos in a lump as we will demonstrate in this section. For this purpose we consider here neutrino number profiles $n_\nu(r)$ with a finite virial radius $\rv$ such that the total number of neutrinos $N_\nu$ is well defined and it is related to $N_*$ and $c$ according to Eq.~(\ref{nonu}). For cosmological purposes we need to compute the average neutrino mass in the lump
\begin{equation}
\langle m_{\nu} \rangle = \frac{4\pi}{N_{\nu}} \int_0^{\rv} dr \, r^2 n_\nu(r) 
m_{\nu}(\phi(r)) \,.
\end{equation}
The gravitational potential of the lump for large distances $r> \rv$ is determined by $\langle m_\nu \rangle$ and $N_\nu$ as
\begin{equation}
\Phi_\nu(r) = -\frac{8\pi}{M^2} \frac{N_\nu \langle m_\nu \rangle}{r} \,.
\end{equation}
The overall size of the cosmologically averaged value of the fluctuations of the neutrino induced gravitational potential determines the size of possibly observable signals for neutrino lumps, as the integrated Sachs-Wolfe (ISW) effect in the CMB anisotropies \cite{Pettorino:2010bv} or the bulk flow of peculiar velocities \cite{Ayaita:2009xm,Ayaita:2009qz}. 

\begin{figure}
\includegraphics[width=8.5cm]{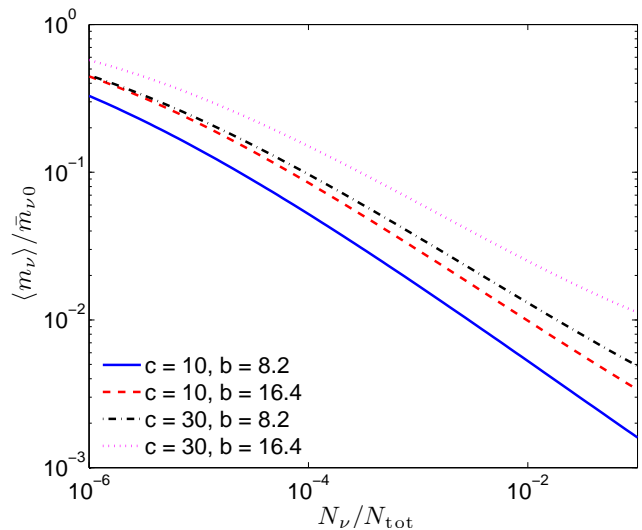}
\caption{\label{figaverage} Dependence of the ratio between the average neutrino mass in a lump and the background value on the number of neutrinos in a lump for a step NFW neutrino profile.}
\end{figure}

In particular, we are interested in the average reduction factor
$r_m(N_\nu)= \langle m_\nu \rangle/\bar{m}_{\nu 0}$ of the neutrino mass within lumps. 
We present in Fig.~\ref{figaverage} the factor $r_m$ as a function of the number of neutrinos $N_\nu$ in the lump. We show our results for  two values of the concentration parameter $c = (10,30)$ 
to which corresponds $N_* = (0.7,0.4) N_{\nu}$ and we assume a self similar relation between lumps such that 
\begin{equation}
\label{rsvector}
r_s = b  \left(\frac{N_\nu}{N_{\rm tot} \, F(c)}\right)^{1/3} {\rm Mpc} \,,
\end{equation}
assuming $b = 8.2$ and $b = 16.4$. This relation arises for a fixed $n_s$  using Eqs.~(\ref{nstar}) and (\ref{nonu}). For a lump with $N_* = 4 \times 10^{-3}$ and $c = 30$, these values of $b$ lead, respectively, to the values of $r_s = 1.3$Mpc and $r_s = 2.6$Mpc that we used in previous sections.
 We find a substantial reduction ($r_m < 1/2$) even for relatively small lumps with 
$N_\nu = 10^{-6} N_{\rm tot}$. The spread of the different lines for the various sets of parameters can be interpreted as an indication of the uncertainty due to our limited knowledge of the actual profile $n_\nu(r)$. The qualitative behavior, however, is common to all cases.

For lumps with number of neutrinos fraction $N_\nu/N_{\rm tot}$ in the range
$10^{-6}$ to $10^{-1}$ the average neutrino mass is between one and three orders of magnitude smaller than its present cosmological value. This result demonstrates the significance of non-linear effects in the growing neutrino scenario, here quantified in terms of the backreaction on the neutrino mass. 

Our result implies important modifications for the gravitational potential of a neutrino lump when compared to linear perturbation theory. Since the neutrino mass is its source on all scales, the magnitude of the gravitational potential is reduced by the same factor $r_m(N_\nu)$ compared to extrapolations based on the cosmological neutrino mass. 

By a similar argument, we expect  an effective suppression of the cosmon mediated attractive force between neutrino lumps which is proportional to $2\beta^2$. This can be estimated
by the averaged interaction strength $\langle \beta\rangle$ for the neutrinos in a neutrino lump, defined as 
\be
\langle \beta \rangle= -M \frac{\partial\ln\langle m_\nu\rangle}{\partial \phi} \,.
\ee
The non-linear effects reduce the attractive force of the lump on a surrounding neutrino fluid by a factor $\langle\beta\rangle/\beta$ as compared to the same number of neutrinos in the absence of non-linear effects. The attraction between two equal lumps is correspondingly reduced by a factor $(\langle\beta\rangle/\beta)^2$. Furthermore, the characteristic time scale for the infall of neutrinos into a lump governed by the fifth force is inversely proportional to the interaction strength. This implies that the characteristic time scale for the infall is increased by a factor $\beta/\langle\beta\rangle$ compared to the consideration excluding non-linear effects and thus results in a slow down of the infall. In the same line, the time scale for the clumping of lumps to larger ones is enhanced by a factor $(\beta/\langle\beta\rangle)^2$.

\begin{figure}
\includegraphics[width=8.5cm]{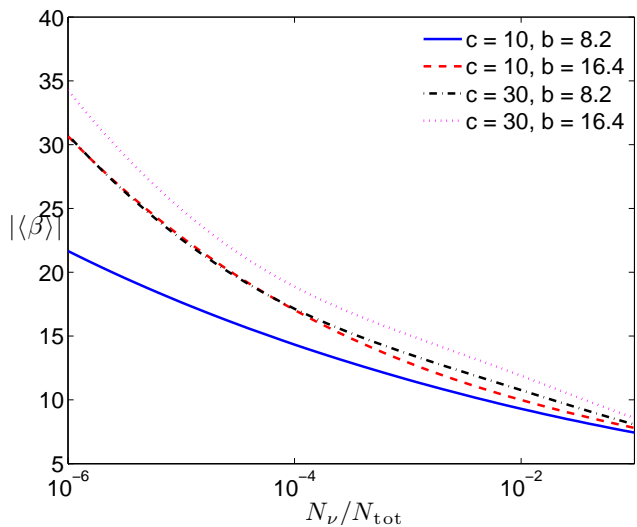}
\caption{\label{effbetafig} The average interaction strength $\langle \beta\rangle$ as a function of the number of neutrinos $N_\nu$ in the lump for a step NFW neutrino profile.}
\end{figure}

In Fig.~\ref{effbetafig}, we show the averaged interaction strength $\langle \beta \rangle$ as a function of the number of neutrinos in the lump for $c=(10,30)$, again assuming $b = 8.2$ and $b = 16.4$. In the range $N_\nu/N_{\rm tot}=10^{-6}$ to $10^{-1}$, we find the averaged interaction strength $ \langle \beta\rangle$ to be reduced by up to one order of magnitude compared to the cosmological value $\beta=-52$ for a step NFW neutrino profile. Again, this suppression is a reflection of the non-linearities in the field equation. We found, however, that within the same radius $R=c \,r_s$, lumps with a pure NFW neutrino profile and with $N_\nu(R)/N_{\rm tot}\simeq 10^{-2}-10^{-1}$ have an averaged interaction strength $\langle \beta \rangle\lesssim 1$, which implies that the effective cosmon-neutrino interaction strength may even be smaller than the gravitational strength. 

In any case, if a large fraction of the neutrinos in the horizon is captured in large neutrino lumps, this may first alter the cosmological field equations for the background fields by order one effects. Second, it may significantly modify the 
growth of neutrino fluctuations with long wavelengths due to backreaction effects from fluctuations with shorter wavelengths. More precisely, owing to neutrino clustering on smaller scales, fluctuations on large length scales will experience a modified effective neutrino mass. In addition, they will be subjected to a smaller effective cosmon force and thus exhibit a substantially reduced growth rate compared to linear extrapolations. Our results for the gravitational potential can be used to compute the impact of neutrino lumps on the large scale structure and the cosmic neutrino background in the form of the Integrated Sachs-Wolfe effect \cite{Pettorino:2010bv}.

In summary, a fluid of neutrino lumps behaves very differently from a fluid of smoothly distributed neutrinos. If the typical size of the lump is sufficiently large and if the true neutrino number distribution in the lump should result in $\langle \beta \rangle \lesssim 1$, then the fluid of lumps behaves quite similarly to dark matter, with only a moderate enhancement of the total attractive force. The time scale for a further growth of structure is comparable to the time scale for 
the gravitational growth of dark matter structures. The early very rapid growth of neutrino structure terminates effectively once the lumps reach a critical size. A detailed estimate of this critical size would be highly interesting as it limits the size of the largest structures that can be expected in our universe.

We also note that the relation between cosmological and terrestrial bounds for the neutrino mass may strongly depend on whether our galaxy is within a large neutrino lump or not. If our galaxy is indeed in a neutrino lump, then the choice of parameters $\beta = -52$ and $\hat{m}$ in Eq.~(\ref{eq:mexp}) (corresponding to a present cosmological neutrino mass $\bar{m}_{\nu 0} = 2.3$eV) would lead to a substantially smaller neutrino mass seen on a terrestrial experiment. Conversely, a present terrestrial bound $m_\nu < 2$eV could be compatible with a substantially larger mass for unbound neutrinos, allowing realistic cosmologies with smaller values of $|\beta|$.

\section{The variation of fundamental parameters}
\label{varalpha}

A scalar field is expected to have couplings with other forms of matter (unless some unknown symmetry principle explicitly forbids them) and consequently, if the field is varying, it generates variations of masses and coupling constants \cite{Wetterich:1987fk,Damour:1994zq,Chiba:2001er,Dvali:2001dd}.
In contrast to the neutrino-cosmon coupling, the interaction between the cosmon and baryons, leptons or photons must be much weaker than gravity in order to avoid conflict with the severe observational bounds. Some suppression mechanism for the field dependence of particle masses (except for neutrinos and possibly dark matter) and dimensionless couplings as gauge or Yukawa couplings  must be at work, perhaps by a fixed point behavior~\cite{Wetterich:2002wm,Wetterich:2008sx}. Unfortunately, the strength of such a suppression is very model dependent such that a variation of a coupling of an observable size is not always guaranteed. However, if the neutrino mass arises from the field dependence of a heavy triplet in the cascade (or seesaw II) mechanism, this directly induces a small field dependence on the Fermi and fine structure constants \cite{Wetterich:2007kr}. Among the various fundamental couplings of the standard model of particle physics, we concentrate here on the fine structure constant $\alpha$. Results for other couplings are similar.

 The possibility that dark energy couples to electromagnetism and that it leads to the variation of the fine structure constant has been raised  numerous times in the literature \cite{Dvali:2001dd,Chiba:2001er}. Claims that the fine structure constant was smaller in the past between redshifts 2--3, $\Delta \alpha/\alpha \sim -5\times 10^{-6}$, suggested by the Keck/HIRES high resolution quasar spectra, have been reported \cite{Murphy:2003hw,Murphy:2003mi}, though these claims are under intense scrutiny and debate \cite{Srianand:2007zz}. On the other hand, geological bounds from the Oklo natural reactor active at an equivalent redshift $z \approx 0.14$ yield $\Delta \alpha/\alpha=(0.7\pm1.8)\times 10^{-8}$ \cite{Gould:2007au} 
and an independent analysis gives
$\Delta \alpha/\alpha =(0.6\pm6.2)\times 10^{-8}$ \cite{Petrov:2005pu}. 
For a linear dependence of $\Delta \alpha/\alpha$ on redshift, we can immediately identify an inconsistency between the Oklo bounds and the claimed cosmological variation at high redshift. This discrepancy is approximately of a factor $10^3/(2/0.14) \approx 70$.
Additionally, strong constraints on the current variation arise from atomic clock measurements,
$\dot{\alpha}/\alpha=(-1.6\pm 2.3)\times 10^{-17}\,{\rm yr}^{-1}$ \cite{rosenband}.
Under the assumption that any observed variation of the fine structure constant of the order $10^{-5}$ at redshifts 2--3 arises from a slowly time varying field which exhibits a small linear coupling to electromagnetism, the atomic clocks and the Oklo  bounds challenge, therefore,  models of dark energy with a monotonic scalar field evolution \cite{Anchordoqui:2003ij,Copeland:2003cv,Bento:2008cn,Dent:2008vd,Nunes:2009dj}. 

There are, however,  several factors  that may alleviate the apparent discrepancy between the bounds on the present variation and the Oklo bounds and the claims of a variation at high redshift. For example, the time evolution of the field may have slowed down in recent times due to the coupling of the quintessence field to matter \cite{Lee:2004vm}, such as in the growing neutrino scenario or in crossover quintessence \cite{Wetterich:2003jt}. We discuss here another effect, namely, that our galaxy may be within a large neutrino lump such that the cosmon field has decoupled from the cosmological evolution at a recent redshift, say $z \approx 1$. We expect then a substantial reduction of the present time variation and predict a different value of the fundamental constants in different regions of the lump and within different lumps. Indeed, we will see that inside a lump the time variation of the cosmon field can be sufficiently small to prevent a violation of the Oklo bound even for the substantial variation of $\alpha$ claimed in Refs.~\cite{Murphy:2003hw,Murphy:2003mi}.

In the following, we assume that we live within a large neutrino lump which formed rapidly around 
$z\simeq 1-2$ and since then it maintained an approximately constant neutrino profile as given in Eq.~(\ref{NFW}) or (\ref{n1}). We will identify a set of model parameters for the growing neutrino scenario, which effectively leads to a decoupling of the value of the scalar field inside a neutrino lump and its cosmological value. 

In the case in which the evolution of the scalar field is sufficiently small, i.e. $|\Delta \phi|  = |\phi-\phi_0| < M$,  we can take the cosmon coupling with electromagnetism to be linear. This can be seen as the first term of a Taylor expansion. The evolution of $\alpha$ is then given by
\begin{equation}
\label{daalinear}
\frac{\Delta\alpha}{\alpha} \equiv \frac{\alpha-\alpha_0}{\alpha_0} = \zeta \frac{\Delta \phi}{M} \,,
\end{equation}
where $\zeta$ is a constant.

Accordingly, the local variation of $\alpha$ (by local we mean as measured with atomic clocks or with Oklo) is given by the difference of the local value of the field at two different epochs ($z\approx 0$ or $z= 0.14$ for atomic clocks and Oklo, respectively). Hence, for Oklo,  we have 
\be
\label{daaoklo}
\left(\frac{\Delta\alpha}{\alpha}\right)_{\rm Oklo} = \frac{\zeta}{M} \left[\phi(z=0.14,r)-\phi(z=0,r)\right] \,,
\ee
where $r$ represents our radial position in the neutrino lump.
Similarly, the cosmological variation of $\alpha$ can be seen as a result of the 
difference between the value of the field at a distance $R_c$ from the center of the lump (where the observed molecular cloud is located)  at high redshift (e.g. $z_c = 2$) before the lump has formed and 
the local value of the field today 
\be
\label{daacloud}
\left(\frac{\Delta\alpha}{\alpha}\right)_{\rm cloud} = \frac{\zeta}{M} \left[\phi(z_c=2,R_c)-\phi(z=0,r)\right] \,.
\ee
For practical purposes we will first identify $\phi(z_c,R_c)$ with the cosmological value $ \bar{\phi}(z_c)$ and we will discuss other scenarios for $\phi(z_c,R_c)$ below. 

In order to solve the discrepancy between the Oklo bound and the claimed variation of $\alpha$ cosmologically, the ratio between Eqs.~(\ref{daaoklo}) and (\ref{daacloud}) 
\be
\label{ratiof}
f(r) \equiv \frac{\phi(z=0.14,r)-\phi(z=0,r)}{\bar{\phi}(z_c=2)-\phi(z=0,r)} \,,
\ee 
must satisfy
\be
-4.4 \times 10^{-3} \lesssim f(r) \lesssim 1.9 \times 10^{-3}  \,,
\ee
where we have used the strongest Oklo bound of Ref.~\cite{Gould:2007au} and assume that $\Delta \alpha/\alpha(z_c=2) = -5.7\times 10^{-6}$ as suggested by the results of Refs.~\cite{Murphy:2003hw,Murphy:2003mi}.

The quantity $f$ above can be thought of as the ratio of two distinct effects. On the one hand we have 
a reduction factor between the local evolution of the field (between redshift $z_l$ of the local observation 
and today) and the background evolution 
\be
f_l(z_l,r) \equiv \frac{\phi(z_l,r)-\phi(z=0,r)}{\bar{\phi}(z_l)-\bar{\phi}(z=0)} \,.
\ee
This quantity is, up to the normalization by the background, identical to   $\Delta \phi$ illustrated in Fig.~\ref{Shifts}.
On the other hand we have a measurement of the 
difference between the value of the field outside the lump in the distant past at redshift $z_c$ and its present value compared to the cosmological evolution in the same time interval
\be
f_c(z_c,r,R_c) \equiv \frac{\phi(z_c,R_c)-\phi(z=0,r)}{\bar{\phi}(z_c)-\bar{\phi}(z=0)} \,.
\ee
Then we obtain that for a generic $z_l$ and $z_c$,  the ratio $f$ is simply given by 
\be
f(z_l,z_c,r) = \frac{f_l(z_l,r)}{f_c(z_c,r,R_c)}\, \frac{\bar{\phi}(z_l)-\bar{\phi}(z=0)}{\bar{\phi}(z_c)-\bar{\phi}(z=0)} \,.
\ee
 Provided the value of the suppression factor  $f_l$ is sufficiently smaller than the 
cosmological shift factor $f_c$, local bounds at $z_l$ become easier to be satisfied even for a large cosmological variation of $\alpha$ at $z_c$. We emphasize that the same ratio $f(r)$ applies to all time variations of couplings to which a linear approximation similar to Eq.~(\ref{daalinear}) holds.

\begin{figure}
\includegraphics[width=8.5cm]{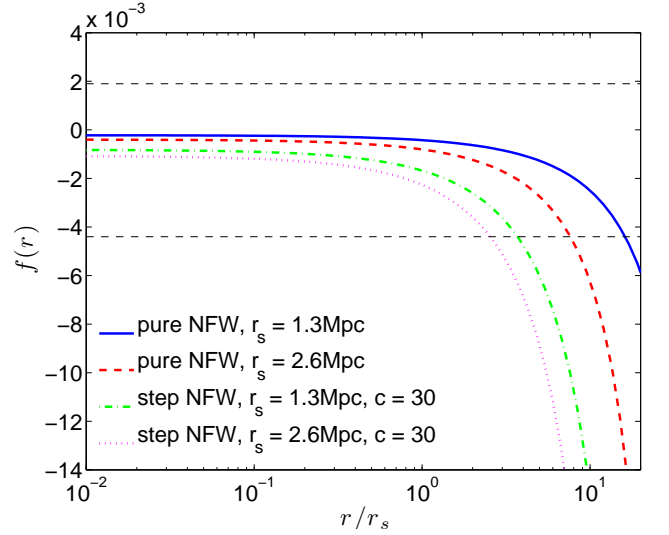}
\caption{\label{fullf} The suppression factor $f(r)$ for four neutrino number profiles with $N_*=4\times10^{-3}$. The horizontal dashed lines represents the values between which the claimed observed cosmological variation of $\alpha$ and the Oklo bounds are consistent.}
\end{figure}

In Fig.~\ref{fullf} we show the overall suppression factor $f(r)$ in Eq.~(\ref{ratiof}) for our model.
We also indicate with the dashed lines the value of $f$ between which the discrepancy between the Oklo bounds and claimed cosmological variation is lifted. In our example with $N_*=4\times10^{-3}$, no discrepancy is present if our local position is  $r \lesssim 10 r_s$ for the pure NFW profile whereas for the step NFW, $r$ is constrained to be $r \lesssim 3 r_s$.

\begin{figure}
\includegraphics[width=8.5cm]{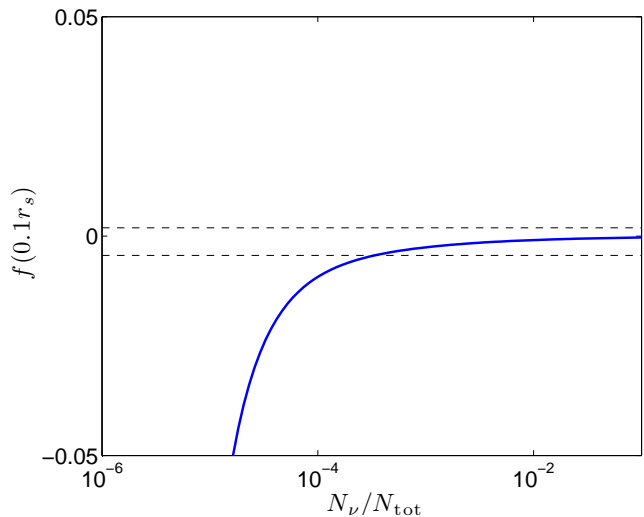}
\caption{\label{fullfNu} The suppression factor $f(r = 0.1r_s)$ for the step NFW profile with respect to the number of neutrinos in a lump for $b = 8.2$ and $c = 30$. The horizontal dashed lines represent the current Oklo bounds.}
\end{figure}

The suppression factor depends strongly on the number of neutrinos in the lump. In Fig.~\ref{fullfNu} we show $f(r = 0.1 r_s)$ as a function of $N_\nu/N_{\rm tot}$ for $b = 8.2$  and $c = 30$. It becomes evident that only for large lumps the claimed observed cosmological variation of $\alpha$ and the Oklo bounds is consistent.

Finally, we notice that the field evolution outside the lump, $\bar{\phi}(z)-\bar{\phi}(0)$, is a useful quantity for comparison, but actually does not appear in the observable quantities. For the local observations, including Oklo, only the difference $\phi(z_l,r)-\phi(0,r)$ matters, while the observable cosmological variations only depend on $\phi(z_c,R_c)-\phi(0,r)$. In this context we note that $\phi(z_c,R_c)$ depends on the observed position of the molecular cloud. If lumps form early enough, it is even conceivable that at a cosmological redshift $z_c \approx 2$, the scalar field is already sufficiently inhomogeneous  such that a possible cosmological time evolution of couplings is replaced by a dependence of $\Delta\alpha/\alpha$ on the position of the molecular cloud. This would result in a dependence of $\Delta\alpha/\alpha$  on the particular observed quasar. In the case of an  inhomogeneous situation,  $\phi(z_c,R_c)$ ought to be replaced by $\phi(z_c,R_c,\theta,\varphi)$, i.e., it depends on the angular coordinates. (For large enough $z_c$ the distance $R_c$ to the cloud can be related to the redshift of the cloud $z_c$.)  A systematic survey of the variation of fundamental couplings with respect to these coordinates would
produce a map of the inhomogeneities of the cosmon field. An analysis suggesting an angular dependence of $\alpha$ has recently been reported \cite{Webb:2010hc}. 

\section{Conclusions}\label{conc}

In this work, we have explored backreaction effects arising in growing neutrino quintessence. They result from the response of local scalar field variations to the formation of large scale non-linear neutrino lumps in the presence of a cosmon-neutrino coupling. The coupling strength is typically much larger than the gravitational interaction. This amplifies the effects of the non-linear nature of the neutrino source term in the field equation inside of a highly-concentrated neutrino lump. By numerically solving for the spatial scalar field profile assuming an NFW distribution for the neutrino number density, we have demonstrated that these non-linearities are manifested in a strong reduction of the neutrino mass within a neutrino lump. Stunningly, for realistic example cases, we found in the interior of a lump, a suppression of the neutrino mass by several orders
of magnitude compared to the cosmological value well outside of the lump. A key result of our work is that the radial profile of the cosmon field and thus of the neutrino mass are effectively frozen inside a lump, once it virializes, while both the cosmological field and the background neutrino mass may continue to evolve with time. The radial dependence of the scalar field and the neutrino mass in the interior of the lump are essentially fixed by the neutrino distribution.

For the time being, substantial uncertainties regarding the neutrino number distribution in the lump subsist. We have investigated here an NFW profile, but it should be mentioned that neutrino lumps differ from dark matter clusters in several important aspects: (i) the relevant time scale of formation is much shorter, such that there is plenty of time to reach a quasi-static equilibrium configuration; (ii) the varying mass of the neutrinos can have important effects for the stable profile; (iii) if the neutrino mass in the core of the lump is very small, the neutrinos with high enough momentum behave close to relativistic particles, leading to a less concentrated profile. The true neutrino profile may be determined by a self-consistent stable static solution similar to Refs.~\cite{Brouzakis:2007aq,Bernardini:2009rc,Bernardini:2009kr}. For this purpose, however,  one needs to gain some insight on the average velocity distribution of the neutrinos inside the lump, resulting in an effective pressure which depends on the radius. The strong reduction of the neutrino mass inside the lump is an effect that seems to be rather robust, independent of the details of the neutrino number distribution. The size of the effect mainly depends on the total number of neutrinos contained in the lump, and to a smaller extent on the core density or concentration. We have demonstrated this by investigating distributions with different neutrino numbers and core densities.     In principle, also the neutrino number distribution can be influenced by a change of the cosmological evolution of $\bar{\phi}$. We have neglected this effect taking the neutrino number distribution to be static after virialization. In view of the effective decoupling of the lump from the cosmological evolution this ansatz is self-consistent, at least as far as the qualitative features are concerned.

We have computed the average neutrino mass $\langle m_\nu \rangle$ within various families of neutrino lumps. For the specific scenario under consideration and for lumps with $N_\nu/N_{\rm tot}$ in the range $10^{-6}$ to $10^{-1}$, we found its value to be between one and three orders of magnitude smaller than the cosmological mass. 
This result reflects the importance of backreaction effects in growing neutrino quintessence. 
In particular,  it causes the gravitational potential of very large neutrino lumps to also be reduced by several orders of magnitude when compared to estimates based on the value of the cosmological neutrino mass. This reduction stems from the fact that the gravitational potential is sourced by the total neutrino mass in the lump and therefore sensitive to the average neutrino mass. Similarly, we found a substantial reduction of the effective coupling between large neutrino lumps and the cosmon.

Finally, we have considered the implications of our findings for the variation of the fine structure constant $\alpha$, which would result from a coupling of the cosmon to electromagnetism, or the variation of other fundamental constants. Variations of particle masses and coupling constants result from variations of the value of the cosmon field $\phi$. We find that the time variation of $\phi$ is substantially reduced inside large neutrino lumps, as compared to the outside variation. If our galaxy is located within such a lump, local time variations -- as observed by precision experiments or the analysis of the Oklo natural reactor -- would be substantially suppressed as compared to cosmological observations. This may weaken the discrepancy between bounds for local variations of $\alpha$ and claims of an observed variation in molecular absorption spectra at high redshifts, which would result from simple linear extrapolations. For sufficiently strong inhomogeneities of the cosmon field at a redshift around $z=2$, the observed values of $\Delta \alpha/\alpha$ could even depend on the position of the molecular clouds and therefore on direction.
In our model, the maximal variation of $\alpha$ is limited by the maximal ratio of the neutrino masses outside and inside our lump. The neutrino mass locally must exceed $m_\nu > 0.05$eV and its value in the background cannot be larger than a few eV without conflicting with cosmological limits. These bounds 
consequently constrain our position inside the lump.

Our investigation illustrates the significant impact of strong backreaction effects in the growing neutrino scenario on the interpretation of cosmological observations. These non-linear effects are expected to be even larger in the case of a field dependent cosmon-neutrino coupling $\beta(\phi)$. Unfortunately, the large size of the backreaction effects also implies that only once backreaction is incorporated properly in the full analysis, the range of parameters of growing neutrino quintessence, which is compatible with cosmological observations, can be analyzed. 
In particular, the growth of large scale neutrino perturbations itself can be expected to be significantly modified. The reason is that after the formation of neutrino lumps on smaller scales these large scale perturbations feel a reduced effective neutrino mass and effective coupling. 

Our findings have important consequences for the effects of neutrino lumps on the cosmic 
microwave background in the form of the Integrated Sachs Wolf effect \cite{Pettorino:2010bv} or on enhanced bulk flows~\cite{Ayaita:2009qz}. Indeed, neutrino lumps would be observable mainly by the gravitational potential induced around them. In this sense, they are very similar to dark 
matter clusters. Galaxies will have tendency to fall into the large scale 
lump potentials. The time when this effect becomes important and the size 
of the effect depends crucially on the size of the gravitational potential 
generated by the lumps and therefore on the effective neutrino mass 
inside the lump.
Also the cosmological background evolution will be modified once a large fraction of the neutrinos is within lumps. 

The growing neutrino scenario seems to be one of the first examples for a realistic backreaction effect which modifies the parameters of the homogeneous and isotropic field equation by order one effects.

\appendix
\section{}

In this appendix, we consider neutrino number density profiles $n_\nu(r)$ with a finite virial radius $\rv$ such that the total number of neutrinos $N_\nu$ is well defined. In this case we can relate the parameters used in the main text to the neutrino number overdensity $\Delta_n\equiv \langle n_\nu\rangle/\bar{n}_{\nu 0}$.

The neutrino structure, bound by the fifth force and by gravity \cite{Wintergerst:2009fh,Brouzakis:2007aq,Bernardini:2009rc,Bernardini:2009kr} has an average density $\langle n_\nu\rangle$ defined by
\be
N_{\nu}=\frac{4\pi}{3}\langle n_\nu\rangle r_{\rm vir}^3 \,.
\label{nav}
\ee
We can rewrite this equation in terms of the ratio of the number of neutrinos in the lump, $N_\nu$, with respect to the total number of neutrinos, $N_{\rm tot}$, within the horizon, 
\be
\rv=\frac{1}{H_0}\left(\frac{N_\nu}{N_{\rm tot}\,\Delta_n}\right)^{1/3},\label{rvirD}
\ee
where $N_{\rm tot}=4\pi\bar{n}_{\nu 0}/(3 H^3_0)$.
The conservation of the number of neutrinos taking part in the formation process of the lump and the self-similarity of the free-fall of collapsing structures may suggest that $\Delta_n$ is a constant. In  this case, however, the ratio of the average energy density $\langle \rho_\nu \rangle$ of the lump to the present neutrino energy density in general does not coincide with $\Delta_n$, $\langle \rho_\nu \rangle/\bar{\rho}_{\nu 0}\neq \Delta_n$. The reason is that with increasing density of the forming lump the mass of the neutrinos in the interior increases slower than the background neutrino mass or it can even freeze or decrease. 
If $\Delta_n$ is a universal constant for all types and sizes of lumps, then it partly fixes the parameters $r_s$ and $n_s$ of the neutrino number density distribution. 
Inserting Eq.~(\ref{rvirD}) into Eq.~(\ref{NFW}), we can determine the functional form of the corresponding NFW distribution for a given $N_\nu$ in terms of $\Delta_n$ and the concentration $c$, 
\bea
n_s&=&\frac{(c\,H_0)^3}{4\pi F(c)}N_{\rm tot}\Delta_n\\
r_s&=&\frac{1}{c\,H_0}\left(\frac{N_\nu}{N_{\rm tot}\,\Delta_n}\right)^{1/3}.
\eea
\begin{figure}
\includegraphics[width=8.5cm]{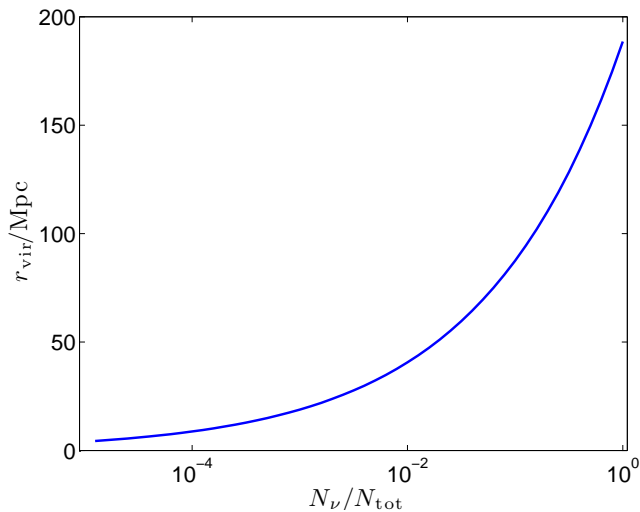}
\caption{\label{R} The virial radius of a neutrino lump as a function of the total number $N_\nu$ of neutrinos it contains for $\Delta_n = 1.1\times 10^{4}$.}
\end{figure}

In Fig.~\ref{R} we plot the resulting functional dependence of the virial radius $\rv$ on the number of neutrinos $N_\nu$ it contains 
for $\Delta_n = 1.1\times 10^4$ according to Eq.~(\ref{rvirD}). For the dedicated numerical simulation in Sec.~\ref{numphi}, the neutrinos contained in a volume with radius $R = 25$Mpc have reached this overdensity at $z = 2.017$. The spherical infall  of the neutrino shells corresponding to radii $r < R$ at $z = 2.017$ is not affected by neutrinos in shells with $r>R$. We may thus discuss a situation with no neutrinos for $r > R$ at $z = 2.017$. The dynamics of the neutrinos in the shells with $r < R$ is the same as in the simulation. For this scenario the value $\Delta_n = 1.1\times 10^{4}$ constitutes a lower bound for the final  overdensity as $\Delta_n$ can only increase due to a still ongoing collapsing process.

\begin{acknowledgments}
NJN is supported by Deutsche Forschungsgemeinschaft, project TRR33 and is also partially supported 
by the projects CERN/FP/109381/2009 and PTDC/FIS/102742/2008. The authors thank Youness Ayaita, Maik Weber and Nico Wintergerst for numerous discussions and are in debt to Nico Wintergerst for also providing the data that allowed the computation of $N_\nu(R)$.
\end{acknowledgments}

\end{document}